\documentclass[doublecol]{epl2} 
\usepackage{amssymb}
\usepackage{amsmath}
\usepackage{graphics}
\usepackage{graphicx}
\usepackage{subfigure}
\usepackage{amsmath,amssymb,graphics,graphicx}
\title{Singularity strength based characterization of financial networks}

\author{Sayantan Ghosh\inst{1,3} \and Uwe Jaekel\inst{4}\and Francesco Petruccione\inst{1,2}}
\shortauthor{Ghosh, Jaekel and Petruccione}

\institute{                    
  \inst{1} Quantum Research Group, School of Physics, University of KwaZulu-Natal, Private Bag X54001, Durban 4000, South Africa\thanks{E-mail:\email{210556397@ukzn.ac.za,say.xen@gmail.com}}\\
  \inst{2} National Institute of Theoretical Physics, University of KwaZulu-Natal, Private Bag X54001, Durban 4000, South Africa\thanks{E-mail:\email{petruccione@ukzn.ac.za}}\\
\inst{3} Dept. of Physical Sciences, Indian Institute of Science Education and Research Kolkata, BCKV Main Campus, P.O. Mohanpur, Dist. Nadia, West Bengal 741 252, India\\
\inst{4} University of Applied Sciences Koblenz, RheinAhrCampus, Dept. of Mathematics and Technology, S\"udallee 2, D-53424 Remagen, Germany\thanks{E-mail:\email{jaekel@rheinahrcampus.de}}
}

\abstract{Financial markets are well known examples of multi-fractal complex systems that have garnered much interest in their characterization through complex network theory. The recent studies have used correlation based distance metrics for defining and analyzing financial networks. In this work the singularity strength is employed to define a distance metric and the existence of hierarchical structure in the Johannesburg Stock Exchange is investigated. The multi-fractal nature of the financial market, which is otherwise hidden in the correlation coefficient based prescriptions, is analyzed through the use of the singularity strength based method. The presence of a super cluster is exhibited in the network which accounts for half of the network size and is homogeneous in the sectoral composition of the South African market.}

\begin{document}

\maketitle
Study and characterization of financial networks has seen a surge in recent years based on the pioneering work by Mantegna \cite{springerlink:10.1007/s100510050929}. Financial networks have been considered in the context of collective behavior \cite{0295-5075-96-4-48004} and asset trees \cite{PhysRevE.68.056110} in the New York Stock Exchange and the Dow Jones Industrial Average \cite{10.1371/journal.pone.0012884}; hierarchical clustering in the Turkish market \cite{springerlink:10.1140/epjb/e2011-20627-6}, the trading volume \cite{Lee2011837} and stock prices \cite{Namaki20113835} in the Korean market. Several methods similar to Mantegna's prescription have also been proposed; like synchronization of chaotic maps \cite{Basalto2005196} and statistical validation based on an unsupervised algorithm \cite{10.1371/journal.pone.0017994} to study financial networks. 
\par
The construction of financial networks requires the identification of a metric that provides the notion of distance between different entities in the network. The metric used in Ref. \cite{springerlink:10.1007/s100510050929}
\begin{equation}
d(i,j)=\sqrt{2(1-C_{ij})},
\end{equation}with $C_{ij}$ being the correlation coefficient is calculated between the return time series of different equities $i$ and $j$ trading in the market. This method however does not account for the well known scaling and multi-fractal nature of financial markets \cite{kantelhardt2002,hes5,hes6,hes7,gs1}.  Hence, to further understand the relation of the multi-fractality and network properties, it would be of considerable interest to employ a multi-fractality based measure of distance and explore the possibility of finding community structure in financial markets.
\par
In this work, we will use the singularity strength or the H\"older exponent to define the ``singularity metric'' that provides the notion of distance between two multi-fractal time series. We calculate the this metric by the use of Wavelet Based Multi Fractal De-trended Fluctuation Analysis (WBMFDFA). This distance metric will then be used to probe the clustering properties of a financial network formed by the equities trading on the Johannesburg Stock Exchange (JSE), through single linkage hierarchical clustering to show the existence of one super-cluster which is homogeneous over the different sectors of the economy. We verify these findings by studying the network properties through the adjacency matrix as a function of a cut-off threshold defined on the basis of the singularity strength. We will also analyze the multi-fractal properties of the JSE by using the Hurst exponent which is related to the multi-fractal scaling exponent $h(r)$ as $H=h(r=2)$. We find that the JSE fluctuations collectively show near-diffusive behavior corresponding to $k^{-2}$ (inverse square law) with a few exceptions which notably are amongst the high capital industries in South Africa.
\par
Mandelbrot in his seminal work in 1963 \cite{mandelbrot1963,mandelbrot1982} proposed the departure from Gaussianity of empirical price change distributions leading to exploration of stationary and non-stationary time series in various fields. The De-trended Fluctuation Analysis (DFA) proposed by Peng \textit{et al.} in 1994 \cite{PhysRevE.49.1685,Ossadnik199464,murad1995estimators,Kantelhardt2001441,PhysRevE.64.011114,PhysRevE.65.041107}) is based on the mono-fractal hypothesis of the time series where the single Hurst exponent \cite{hurst1951} is sufficient to explain the self similarity of the time series. Kantelhardt \textit{et al.} subsequently proposed the generalized Multi Fractal De-trended Fluctuation Analysis (MFDFA) \cite{kantelhardt2002} in 2002, which has found wide applications in understanding the scaling behavior of financial markets (for example see \cite{hes5,hes6,hes7,gs1}). Using the multi-resolution capability of wavelets \cite{farge1,torrence1}, Manimaran \textit{et al.} proposed the Wavelet Based MFDFA (WBMFDFA) \cite{mani1,mani2,mani3} in 2005 which has been used to study the emergence of scaling and self-similar behavior in various financial markets \cite{Ghosh20114304}.This method has also been applied to other areas for example, quark-hadron phase transition in plasma data \cite{Manimaran20103703} and cancer detection \cite{Ghosh:11}. 
\par
In one dimensional Discrete Wavelet Transform (DWT) \cite{farge1,daubechies1992}, a real valued function $\xi(\vec{x})$ can be decomposed as
\begin{equation}
\xi(\vec{x})=\sum\limits_{i=-\infty}^{\infty}a_i\phi_i(\vec{x})+\sum\limits_{i=-\infty}^{\infty}\sum\limits_{j\geq0}b_{ij}\psi_{ij}(\vec{x})
\label{eq:eqDWT}
\end{equation}
where, $\phi(\vec{x})$ and $\Psi(\vec{x})$ are square integrable functions forming the orthonormal basis for a $L^2(\mathbb{C})$ Hilbert space and are called the father and mother wavelets respectively. They are subject to the admissibility conditions: $\int \phi(\vec{x})d\vec{x} < \infty$, $\int \Psi(\vec{x})d\vec{x}=0$, $\int \phi(\vec{x})^*\Psi(\vec{x})d\vec{x}=0$ and $\int \vert \phi(\vec{x})\vert^2 d\vec{x}=\int \vert \Psi(\vec{x})\vert^2 d\vec{x}=1$. The $\psi_{ij}(\vec{x})$ are called the daughter wavelets and are related to the mother wavelet $\Psi(\vec{x})$ by scaling and translation (and in higher dimensional cases also rotation) by $\psi_{ij}(\vec{x})=2^{i/2}\Psi(2^i\vec{x}-j)$, where, $i$ and $j$ are the scaling and translation parameters respectively. Thus, at the $i$th scale, the daughter wavelet $\psi_{ij}(\vec{x})$ is $2^i$ times the height and $2^{i/2}$ the width of the  mother wavelet $\Psi(\vec{x})$. In eq. (\ref{eq:eqDWT}), $a_i$ and $b_{ij}$ are the low pass and high pass coefficients which respectively carry information about the low frequency behavior or ``average behavior'' and high frequency behavior or ``detail behavior'' of the function. In case of the Daubechies wavelets, the $\Psi(\vec{x})$ is also made to satisfy the vanishing moment condition, that is, $\int \Psi(\vec{x})\vec{x}^p d\vec{x}=0, p \leq P$; the analysis of the $P$th order derivative of the function $\xi(\vec{x})$ becomes possible. This is instrumental in the extraction of trends of various polynomial orders from the function. For example, since the second wavelet of the Daubechies family Db-4 has two vanishing moments, it can be used to extract constant and monomial trends, while Db-6 with three vanishing moments can extract constant, linear and quadratic trends. This property is particularly helpful in performing a fluctuation analysis where we are interested in specific scales and specific patterns of the function.
\par
In WBMFDFA, given a non-stationary time series $x(t)$, we firstly make it stationary by evaluating the normalized log return series \cite{mandelbrot1963} given by $R(t)=\frac{r(t)-\langle r(t)\rangle}{\sigma_{r(t)}}$, where $\langle \cdots \rangle$ represents the time average, $r(t)=\log x(t+1)-\log x(t)$ is the log returns and $\sigma_{r(t)}=\sqrt{\langle r(t)^2\rangle-\langle r(t)\rangle^2}$ is the volatility of the series. We then perform the fluctuation extraction on the profile calculated as $Y(n)=\sum\limits_{t=1}^{n}R(t),n\in[1,N-1]$, where $N$ is the length of the return series. The fluctuations at every scale are obtained by reconstructing the trend from the low pass coefficients obtained by performing a DWT with the Db-4 wavelet and then subtracting it from the $Y(n)$. The details of the method have been discussed in \cite{mani1,mani2,mani3,Ghosh20114304,Ghosh:11}. Since the asymmetry of the Db-4 basis leads to convolution errors and edge effects \cite{farge1,torrence1}, we reverse the profile, perform the same fluctuation extraction on the reversed profile and take the average to avoid the artificial errors that may be produced and obtain the fluctuation $f_j$ at each scale $j$. These fluctuations are then segmented into $M_w=\mbox{int}(N/w)$ non-overlapping segments, where $w$ is the window size. The $w$ is related to the scale $j$ by the number of filter coefficients for the given wavelet. The resulting $r$th order fluctuation function $F_r(w)$ ($r\in \mathbb{Z}$) is obtained \cite{kantelhardt2002} as
\begin{equation}
F_r(w)\equiv\left(\frac{1}{2M_w}\sum\limits^{2M_w}_{m=1}\left[F^2(m,w)\right]^{r/2}\right)^{1/r}.
\end{equation}
It can be immediately observed that $F_r(w)$ diverges at $r=0$ and hence $F_0(w)$ is calculated as
\begin{equation}
F_0(w)=\exp\left[\left(\frac{1}{2M_w}\sum\limits^{2M_w}_{m=1}\log\left[F^2(m,w)\right]^{r/2}\right)^{1/r}\right].
\end{equation}
For a self similar process, the fluctuation function $F_r(w)$ follows a scaling law, the scaling function $h(r)$ is given by $F_r(s)\sim w^{h(r)}$ which is related to the classical partition function based multi-fractal scaling exponent $\tau(r)$ by $\tau(r)=rh(r)-1$ \cite{stanley1988,eke2002}. This $\tau(r)$ is related to the singularity spectrum $f(\beta)$ which represents the dimension of the subset of the time series, by a Legendre transform $\beta=d/dr [\tau(r)]$ where $\beta$ is called the singularity strength or the H\"older exponent and $f(\beta)=r\beta-\tau(r)$ \cite{kantelhardt2002}. Thus, the singularity strength, singularity spectrum and the scaling function are related by $f(\beta)=r\{\beta-h(r)\}+1$.
\par
The singularity strength converges to zero in the case of mono-fractals and increases with multi-fractality. We must however, note that though the Hurst exponent defined as $H\equiv h(r=2)$, is bounded in $[0,1]$; the singularity strength only has a lower bound of zero for the mono-fractal case and higher the multi-fractality, more is the width of the singularity spectrum. This makes the width defined by \begin{equation}\gamma=\textrm{max}(\beta)-\textrm{min}(\beta)\end{equation} a perfect candidate for characterizing a multi-fractal time series also allowing for the introduction of a metric for comparing different time-series. We must note here that $H=0$ implies a white noise $(f^{-1})$, $H=0.5$ corresponds to a $f^{-2}$ power law process and $H=1$ is the signature of inverse cubic law process $f^{-3}$. Processes exhibiting $H=0.5$ and $H=1$ are also called as diffusive or Brownian and ballistic processes respectively.
\par
We define the ``singularity metric'' for two multi-fractal time series $\{X:x_1,x_2,x_3,\ldots,x_n;n\in\mathbb{Z},x_i\in\mathbb{R}\}$ and $\{Y:y_1,y_2,y_3,\ldots,y_n;n\in\mathbb{Z},y_i\in\mathbb{R}\}$ as $\rho:\mathbb{R}\times \mathbb{R}\rightarrow \mathbb{R}^+$, where, $\mathbb{R}^+\equiv[0,\infty)$ such that
\begin{equation}
\rho(X,Y)=\vert\gamma_X-\gamma_Y\vert.
\label{eq:rho}
\end{equation}
It can be easily verified that this singularity metric $\rho$ satisfies the conditions: $\rho(X,Y)\geq 0$ (non-negativity), $\rho(X,Y)=0 \Leftrightarrow \gamma_X=\gamma_Y$, $\rho(X,Y)=\rho(Y,X)$(symmetry) and $\rho(X,Z)\leq \rho(X,Y)+\rho(Y,Z)$ (subadditivity). Thus, for given $n$ time series, we can construct the matrix $\rho_{n\times n}$ as
\begin{equation}
\rho_{ij}=\vert\gamma_i-\gamma_j\vert.
\label{eq:eqsing}
\end{equation}
\par
In this work, we use the historical closing prices of 580 of the approximately 1200 equities listed under 472 companies trading on the Johannesburg Stock Exchange (JSE) from 1$^{st}$ January, 1990 to 29$^{th}$ August, 2008 sampled at one trading day interval. It is particularly interesting to analyze the JSE as, a) it is a developing economy and b) since 2001, it has through an agreement allowed cross trading with the London Stock Exchange (FTSE) which implies that the JSE will possibly be affected by the European financial weather. The South African economy can be classified into five major sectors: natural resources, agriculture, manufacturing, services industry and trade \& investment. Of these, natural resources and the services industry are the most dominant sectors. 
\par
The graph partitioning methods of finding $k$-clusters in large graphs are shown to be unreliable due to the lack of a-priori knowledge of the exact number of clusters in the graph and thus failure to correctly partition the nodes into clusters \cite{polya1998,andrews1976,Santo2010}. Hence, multilevel clustering techniques like agglomerative clustering algorithms and divisive clustering algorithms are used in order to reveal the hierarchical structure of the networks. An agglomerative clustering algorithm is a bottoms-up approach where, we start off as a set isolated nodes and as edges are added, based on their similarity, clusters are formed. It has been particularly useful in detecting hierarchical structures in large networks from social and biological networks (for example see \cite{Girvan2002,Newman2006}) and in financial networks (for example see\cite{Arenas200893,pan2008}). Hence, using the singularity metric $\rho$, we define the distance between each equity and use the single linkage agglomerative hierarchical clustering technique to organize the equities into clusters. 
\par
\begin{figure}[h]
\centering
\includegraphics[width=0.4\textwidth,height=0.3\textheight]{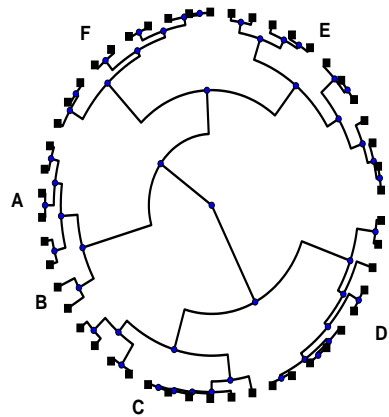}
\caption{\label{fig:dend}Dendrogram representing the hierarchical clustering of the equities using the $\rho$ defined in eq. (\ref{eq:rho}). For the sake of illustrative convenience, we have showed the dendrogram only to the top six branches. Notice the clusters A, B, C, D, E and F which contain 51.73\%, 21.97\%, 12.98\%, 3.81\%, 4.94\% and 8.65\% of the total leaves respectively.}
\end{figure}

As shown in the dendrogram in fig. \ref{fig:dend}, 
the network breaks down into six dominant clusters denoted by A, B, C, D, E and F; each containing 51.73\%, 21.97\%, 12.98\%, 3.81\%, 4.94\% and 8.65\% of the total leaves respectively. We must note here that we have shown only the top six branches for the sake of brevity. This implies the presence of a super-cluster A that contains over half of the nodes. However, to validate this finding, let us look towards finding the graph properties of the network.
\par
\begin{figure}[h]
\includegraphics[width=0.4\textwidth,height=0.3\textheight]{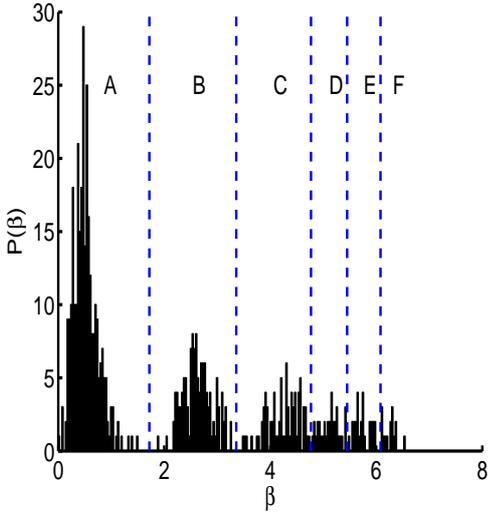}
\caption{\label{fig:dbhist}The probability distribution $P(\rho_{ij})$ of the singularity metric elements $\rho_{ij}$. The multi-modal distribution is evident which allows us to make segmentation of the distribution function into six distinct groups and perform a network analysis with them.}
\end{figure}

Looking at the distribution of $\beta$, as seen from fig. \ref{fig:dbhist}, 
we can see that the $\beta$ shows a multi-modal density distribution. Hence, can collect the nodes falling under a specific range such that A$\in[$min$(\rho),1.72)$, B$\in[1.72,3.36)$, C$\in[3.36,4.77)$, D$\in[4.77,5.45)$, E$\in[5.45,6.08)$ and F$\in[6.08,$max$(\rho)]$. This provides us information about the percentage of nodes falling under a specific category. We have the following distribution of nodes in each of the six categories from A--F: A=51.90\%, B=21.97\%, C=13.15\%, D=5.88\%, E=4.67\%  and F=2.42\%. We can see that 51.90\% of the nodes are very close to each other, while 2.42\% are the farthest from each other. This is also consistent with our findings from the hierarchical clustering. The three biggest clusters have a comparable number of nodes. 
\par
Given a undirected graph $\mathcal{G}=(\mathcal{V,\mathcal{E}})$ with no self loops, where $\mathcal{V}$ is the set of all vertices or nodes and $\mathcal{E}$ is the set of all edges in the graph $\mathcal{G}$, it is represented by its adjacency matrix $\mathcal{A}$ such that $\mathcal{A}_{ij}=1$ if the nodes $i$ and $j$ are connected and $0$ otherwise. Also $\mathcal{A}_{ii}$ is zero since the graph has no self loops. The average degree of the graph $\mathcal{G}$ is defined as $\langle k\rangle=1/(N(N-1))\sum\limits_{i,j,i\neq j}\mathcal{A}_{ij}$ \cite{Boccaletti2006175}. The characteristic path length which is the mean of all the geodesics between the nodes is given as $L=1/(N(N-1))\sum\limits_{i,j,i\neq j}d_{ij}$ where, $d_{ij} \in \mathcal{D}$ is the geodesics between the nodes $i$ and $j$ and $\mathcal{D}$ is the distance matrix. The graph efficiency is calculated as $E=1/(N(N-1))\sum\limits_{i,j,i\neq j}1/d_{ij}$. The average clustering coefficient for a graph  is given by 
\begin{equation}
\langle C\rangle =\frac{1}{N}\sum\limits_i \frac{2 \{\vert e_{jk}\}\vert}{k_i(k_i-1)}: v_j,v_k\in \mathcal{V}_i, e_{jk}\in \mathcal{E}.
\end{equation}
and the vertex centrality, which is a score of how important is a node in a given topological configuration for the node $i$ is given by
\begin{equation}
B_i=\sum\limits_{j,k,j\neq k}\frac{n_{jk}(i)}{n_{jk}},
\end{equation}
where, $n_{jk}$ is the number of geodesics between the nodes $j$ and $k$ and $n_{jk}(i)$ are the number of geodesics between $j$ and $k$ passing through $i$. The average vertex centrality is given by $B=\sum\limits_{i}B_i$. We must note that these quantities are also dependent on the topology of the graph under consideration. The computation of these quantities was done using the MATLAB BGL software provided in \cite{gleich2009-thesis}.
\par
\begin{figure}
\includegraphics[width=0.475\textwidth,height=0.3\textheight]{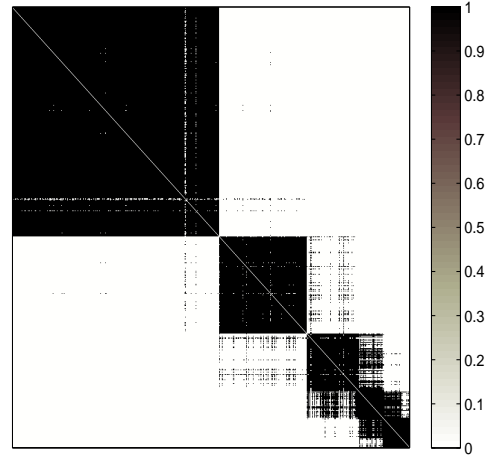}
\caption{\label{fig:adj}The adjacency matrix at $A_{ij}(\xi^*)$, the critical value of $\xi=0.96$. Notice the presence of overlapping communities in the region corresponding to $\rho_{ij}>4.47$. Also, we can see that the communities only interact with their neighbors.}
\end{figure}
\begin{figure}[h]
\centering
\includegraphics[width=0.45\textwidth,height=0.3\textheight]{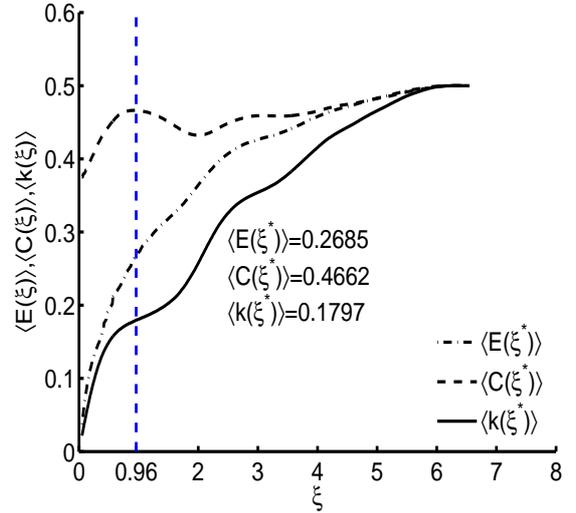}
\caption{\label{fig:clust}Graph Efficiency $\langle E(\xi) \rangle$, clustering coefficient $\langle C(\xi)\rangle$ and average graph degree $\langle k(\xi)\rangle$ plotted as a function of $\xi$ for the network $\mathcal{A}(\xi)$. Notice that at $\xi^*=0.96$, the clustering coefficient is $0.4662$ and falls off in the region $[0.96,2]$ to rise again with a gentler slope. Correspondingly, in the same region, the average graph degree rises with a very small slope. However, there is no change in the slope of the graph efficiency. At $\xi^*$, $\langle E(\xi^*)\rangle$ and $\langle k(\xi^*)\rangle$ are $0.2685$ and $0.1797$ respectively.}
\end{figure}

We define a threshold parameter $\xi \in [$min$(\beta),$max$(\beta)]$ such that, we can calculate the adjacency matrix $\mathcal{A}(\xi)\equiv \mathcal{A}_{ij}(\xi)=1$ when $\rho_{ij}<\xi$ and zero otherwise. For example, we have plotted the adjacency matrix formed for $\xi=0.96$ in fig. \ref{fig:adj}.
We get the corresponding quantities average degree, characteristic path length, graph efficiency and average clustering coefficient $\langle k(\xi)\rangle$, $L(\xi)$, $E(\xi)$ and $\langle C(\xi) \rangle$ as functions of the threshold parameter $\xi$ respectively. We have plotted some of these quantities as a function of the threshold parameter $\xi$ in fig. \ref{fig:clust} for $\mathcal{A}(\xi)$. 
\par
We can observe from the definitions of the graph properties that more nodes are added to the adjacency matrix as we go over the range of $\xi$; and hence $\langle k(\xi)\rangle$, $\langle C(\xi)\rangle$ and $\langle E(\xi)\rangle$ should be increasing functions of $\xi$. However, from fig. \ref{fig:clust}, we observe that the average clustering coefficient $\langle C(\xi)\rangle$ decreases in the interval $[0.96,2]$ and then rises again. Correspondingly, the average degree $\langle k(\xi) \rangle$ also shows a change of slope in this region. It is interesting to note that the quantity $\langle k(\xi) \rangle$ changes very slowly in this region, while the graph efficiency $\langle E(\xi)\rangle$ does not show much deviation from the expected monotonic rising behavior. This is an interesting finding since it implies that though the  degree of the graph does not change appreciably, the graph undergoes substantial ``re-wiring'' in this regime which indicates dynamical restructuring of the relationship between the various equities when in certain phases of their multi-fractality. This could imply the occurrence of a structural phase-transition since the $\langle C(\xi)\rangle$ increases rapidly at low $\xi$, reaches a peak value at $\xi=0.96\equiv\xi^{*}$ and falls off to rise again but with a gentler slope. This value of $\xi^{*}$ can be considered as the critical value for the network for which $\langle C(\xi^*)\rangle=0.4662$. The relationship of high clustering coefficient and modularity has recently been discussed in \cite{barmpoutis2010networks}. 
 
\begin{figure}[h]
\centering
\includegraphics[width=0.45\textwidth,height=0.3\textheight]{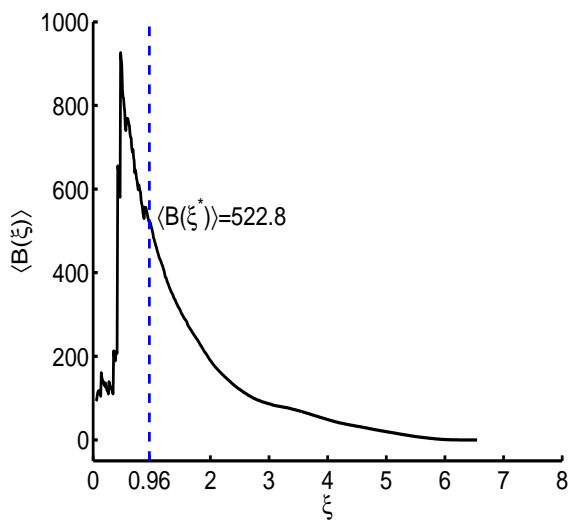}
\caption{\label{fig:betcent}The unnormalized average betweeness centrality $\langle B(\xi)\rangle$ plotted as a function of the cut-of threshold $\xi$. Note that though at $\xi^*$, there is a local maxima, the global maxima lies at $\xi=0.5366$, indicating an optimal network configuration at this value.}
\end{figure}
\par
In fig. \ref{fig:betcent}, we have plotted the unnormalized $\langle B(\xi)\rangle$ as a function of $\xi$. It can be seen that the $\langle B(\xi)\rangle$ reaches a peak value of $926.19$ at $\xi=0.5366$, while at $\xi^*$ it is significantly lower at $522.8$. However, it is also evident that there is a local maxima of $\langle B(\xi)\rangle$ at $\xi^*$. This could possibly mean that the most optimal configuration of the network is at $\xi=0.5633$ which implies a narrow band of the singularity strength where the equities differ from each other by $\rho\sim 0.5633$. This also corresponds to the maxima of the distribution function followed by $\beta$ as is evident from fig. \ref{fig:dbhist}.
\begin{figure}
\centering
\includegraphics[width=0.47\textwidth,height=0.3\textheight]{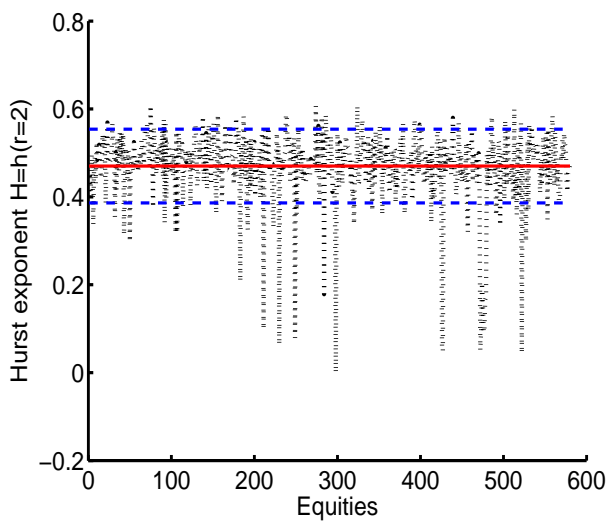}
\caption{\label{fig:hurst} Hurst exponent $H\equiv h(r=2)$, plotted for the 578 equities analyzed in this work. The mean $\langle H\rangle$, represented by the red line is at $0.4701$ and the standard deviation is represented by the blue broken lines at $0.4701 \pm 0.0840$ respectively. We can see that the some of the equities deviate significantly from the mean and the equity Kruger Rand -- Half at $x=296$ (on the $x$-axis has) $H(296)=0$.}
\end{figure}
\par
Figure \ref{fig:hurst} represents the Hurst exponent calculated at h(r=2) for the different equities trading on the JSE. We observe that the mean Hurst exponent for the equities is $H=0.4701\pm0.0840$ which indicates near-diffusive behavior. It can also be seen that 90 equities deviate significantly from the trend. However, only one equity namely Kruger Rand - Half (ticker symbol KRHT) shows $H=0$. Amongst the deviating equities metal, mining, financial institutions services sector and holding companies are the major contributors, notably the platinum mining companies like Anglo American Platinum, Aquarius Platinum, Eastern Platinum, Impala Platinum and Highveld Steel \& Vanadium. Some other notable companies in this range are the AECI, Banro Corp., British American Tobacco, ABSA Bank, Alcoa Steel, ArcelorMittal, BHP Billiton, Goldone Internatinal, First Rand Limited (parent company of the First National Bank), Good Hope Diamond Mines, the Anglo American Group and Amalgamated Breweries Ltd., a subsidiary of the SABMiller Plc. which is one of the largest breweries in the world.
\par
In conclusion, in this work, we have explored financial networks from a multi-fractal analysis point of view. The resulting study throws light on the fact that there exists a hierarchical community structure in the South African financial market based on the multi-fractality of the individual equities. The fact that the singularity strength can be used to find such a clustering can be instrumental in applying this technique to other kinds of time series. Also, we were able to show the presence of one super-cluster, through both agglomerative hierarchical clustering and forming a graph based on the singularity metric. We also showed the presence of a narrow band-width difference of H\"older exponent for which the network shows maximum vertex centrality. The identification of a critical threshold parameter value $\xi^*$ which is almost double the value of $\xi$ for a optimum network configuration is indicative of a dynamic topological changes in the network in a particular regime of multi-fractality. We have also explored the existence of a collective sub-diffusive motion of the equities determined by the mean Hurst parameter of $0.4701\pm0.0840$. This also indicates the presence of a power law behavior of the fluctuations. However, the deviation of a select few high capital industries from the average behavior is interesting which could possibly be due to turbulent geo-socio-political factors in the country in the last few decades and which would have to be checked with respect to the other economies and the socio-political factors modeled in conjunction with wider and more sophisticated studies. 
\acknowledgements
SG would like to thank the CQT, UKZN, Durban for the financial support given to this research. He would also like to acknowledge Jalpa Soni and Prasanta K. Panigrahi for stimulating discussions. The authors also thank Rosario N. Mantegna for useful insights. UJ and FP  gratefully acknowledge  support by the National Research Foundation and the Deutsche Forschungsgemeinschaft.This work is based upon research supported by the South African Research Chair Initiative of the Department of Science and Technology and National Research Foundation. The authors would like to thank Michael Chadbourne of Investec Bank for providing access to the financial data.
\bibliographystyle{eplbib}
\bibliography{draft2.bbl}
%

\end{document}